\documentclass[12pt]{article}
\usepackage{amsmath,amssymb,bbm}
\def\Box{{\hbox{$\sqcup$}\llap{\hbox{$\sqcap$}}}}
\renewcommand{\theequation}{1.\arabic{equation}}

\begin{document}

\begin{flushright}
hep-th/10071290
\end{flushright}

\vskip 2cm
\begin{center}
{\Large\bf NS-branes in 5d brane world models}\\
\end{center}

\vskip 1cm
\begin{center}
{\bf Eun Kyung Park\footnote{E-mail : ekpark@ks.ac.kr}
and Pyung Seong Kwon\footnote{E-mail : bskwon@ks.ac.kr}\\
\vskip 0.6cm}
\end{center}

\begin{center}
{$^1$Basic science research center, Kyungsung University,\\
Pusan 608-736, Korea \\

\vskip 0.3cm

$^2$Department of Physics, Kyungsung University,\\
Pusan 608-736, Korea}
\end{center}

\thispagestyle{empty}

\vskip 2.0cm
\begin{abstract}
We study codimension-1 brane solutions of the 5d brane world models compactified on $S_1 / \mathbb{Z}_2$. In string theoretical setup they suggest that the background branes located at orbifold fixed points should be NS-branes (in the five dimensional sense), rather than D-branes. Indeed, the existence of the background NS-branes is indispensable to obtain flat geometry $M_4 \times S_1 / \mathbb{Z}_2$ where $M_4$ represents the 4d Minkowski spacetime, and without these branes the 5d metric becomes singular everywhere. This result is very reminiscent of the $(p+3)$d effective string theory \cite{1} where the NS-NS type $p$-brane is indispensable to obtain a flat geometry $R_2$ or $R_2 /\mathbb{Z}_n$ on the transverse dimensions. Without this NS-NS type $p$-brane the 2d transverse space becomes a pin-shaped singular space. The correspondence between these two theories leads us to a conjecture that the whole flat backgrounds of the string theory inherently involve the NS-branes implicitly in their ansatz, and hence the true background $p$-branes immanent in our spacetime may be NS-branes, instead of D-branes. We argue that this result can have a significant consequence in the context of the cosmological constant problem.
\end{abstract}

\vskip 1.6cm
\medskip
\begin{center}
\vskip 1cm
\end{center}

\newpage
\setcounter{page}{1}
\setcounter{footnote}{0}

\baselineskip 6.0mm

\hspace{-0.7cm}{\bf 1. Introduction}
\vskip 0.5cm
\hspace{-0.6cm}In the usual brane world scenarios it is basically assumed that our universe is a stack of D3-branes (or D$p$-branes wrapped on a ($p-3$)d compact space) with standard model (SM) fields living on it [2-8]. But recently there was an argument \cite{1} that true background $p$-brane immanent in our spacetime may be NS-NS type brane, rather than D-brane. In \cite{1} the authors showed that in ($p+3$)d string theory the presence of NS-NS type $p$-brane with negative tension is indispensable to obtain background $R_2$ or $R_2 /\mathbb{Z}_n$ on the transverse dimensions, and the usual codimension-2 brane solutions with these background geometries already involve the negative tension NS-brane implicitly in their ansatz.

In this line of studies it was also argued in \cite{9} that the NS-brane plays an important role in the context of the cosmological constant problem in the ($p+3$)d brane world models where the NS-brane appears as a background brane on which the SM-brane is to be set. Indeed, brane world models with NS-branes have been already considered in the literature [10-13] including "Little String Theory"(LST) [14-16]. But the main object of these papers is to obtain SM- or Yang-Mills gauge theories irrespectively of the cosmological constant problem. In \cite{9}, it was shown that the bulk geometry, as well as the flat intrinsic geometry of the brane, is practically insensitive to the quantum fluctuations of SM-fields with support on the SM-brane in the presence of the background NS-brane, which then leads to a new type of self-tuning mechanism with which to solve the cosmological constant problem.

Apart from this, 5d brane world models, as being the simplest higher dimensional models, have been extensively studied in the hope that they may provide the first step toward the solution of the hierarchy problem \cite{8,17,18} and the cosmological constant problem as well [19-23]. Among these models, the one proposed by Randall and Sundrum is of particular interest in the context of the hierarchy problem. In this Randall-Sundrum (RS) model \cite{17} the desired $TeV$ physical mass scale can be obtained from the fundamental Planck scale $\sim 10^{19} GeV$ through an exponential hierarchy generated by an exponential warp factor. But at the same time, the exponential warp factor causes the problems associated with proton decay and neutrino mass etc.\cite{24}, though they can be circumvented by assuming that the standard model fermions and gauge fields do not localized on the IR(visible) brane \cite{25}.

The RS-model can be regarded as a dimensional reduction of the eleven-dimensional theory \cite{26}, and in this sense it is not directly associated with the 10d superstring theory. But one can consider different type of 5d models which are directly obtained from the string theoretical setup, but still compactified on $S_1 / \mathbb{Z}_2$ as in the RS model. In these models the warp factor reduces to a constant and the 5d metric is therefore factorizable. Besides this, it turns out that the background branes located at the orbifold fixed points are NS-NS type branes, instead of D-branes, which is quite unexpected because our spacetime is generally believed to be a stack of D-branes in the usual brane world scenarios. In the 5d brane world models the 3-brane is a codimension-1 brane, and the solutions of these models usually take completely different forms as compared with those of the ($p+3$)d brane world models where the $p$-brane appears as a codimension-2 brane. In this paper we will first briefly examine the codimension-1 brane solutions of the string-inspired 5d gravity-scalar theory to show that the existence of NS-branes is indispensible to obtain the flat geometry $M_4 \times S_1 / \mathbb{Z}_2$, and without these branes the 5d metric becomes singular everywhere.

This result is very reminiscent of the $(p+3)$d effective string theory \cite{1} where the NS-NS type $p$-brane is indispensible to obtain flat geometry $R_2$ or $R_2 / \mathbb{Z}_2$ on the transverse dimensions, and without this brane the 2d transverse space becomes a pin-shaped singular space. The correspondence between these two theories leads us to suspect that the whole flat background geometries of the string theory inherently involve the NS-branes implicitly in their ansatz, and hence the background $p$-branes immanent in our spacetime may be NS-NS type branes, instead of D-branes. In this paper it is also argued that this result can have a significant consequence in the context of the cosmological constant problem.

\vskip 1cm
\hspace{-0.7cm} {\bf 2. 5d action and field equations}
\renewcommand{\theequation}{2.\arabic{equation}}
\vskip 0.5cm
\hspace{-0.6cm}We begin with a 5d action \cite{22}
\begin{equation}
I_5 = \frac{1}{2 \kappa^2} \int d^5 x \sqrt{-G}\, \big[R - k (\nabla \phi)^2 - e^{b \phi} \lambda \big]\,\,,
\end{equation}
plus a brane action
\begin{equation}
I_{\rm brane} = - \sum_{i} \int d^4 x \sqrt{-g}\,\, T^{(i)} (\phi)\Big|_{y=y_i} \,\,,
\end{equation}
where $y_{i}$ represents the position of the $i$-th brane in the fifth coordinate, while $T^{(i)} (\phi)$ the tension of the $i$-th brane coupled with a scalar $\phi$. In most cases of our discussion $T^{(i)} (\phi)$ takes the tree level form
\begin{equation}
T^{(i)} (\phi) = T^{(i)}_0 e^{\alpha \phi} \,\,, ~~~~~~~~(T^{(i)}_0 = {\rm const.})\,\,.
\end{equation}
In the RS-model, $T^{(i)} (\phi)$ is given by (2.3) with $\alpha=0$ because the RS-model does not contain scalar field. In the tree level string theory, on the other hand, $\phi$ is identified with the dilaton and $T^{(i)} (\phi)$ is also given by (2.3), but this time $\alpha$ is assigned to be either $\frac{2}{3}$ or $\frac{5}{3}$ depending on which brane is considered in the theory. If the brane is NS-brane $\alpha$ is $\alpha=\frac{2}{3}$, which correspond to the factor $e^{-2 \phi}$ in the string frame, while for the D-brane $\alpha$ is $\alpha=\frac{5}{3}$, which corresponds to the factor $e^{-\phi}$ in the string frame.

For the background geometry we assume that the 5d spacetime is compactified on $S_1 / \mathbb{Z}_2$, and two 3-branes are located at the orbifold fixed points $y=0$ and $y = \pi R_c$ respectively, where $R_c$ is the radius of $S_1$. Once we choose $T^{(i)}(\phi)$ as in (2.3) the boundary conditions at the orbifold fixed points require that $\alpha ={b}/{2}$ (we will show this later), and the action (2.1) plus (2.2) describes a family of 5d theories parameterized by $k$, $b$, and $\lambda$. Among these 5d theories we are particularly interested in the case $k=b=\frac{4}{3}$, which corresponds to a 5d effective string theory\footnote{The action (1) may be identified as a truncation of the type IIB superstring compactified on 5-torus \cite{27}.} represented in the Einstein frame. In this case the condition $\alpha ={b}/{2}$ implies $\alpha = \frac{2}{3}$, and the branes located at the orbifold fixed points are identified as NS-branes (not D-branes!).

The field equations arising from (2.1) plus (2.2) are the Einstein equation
\begin{eqnarray}
\sqrt{-G} \,\,\big[\, R_{AB} - \frac{1}{2} G_{AB} R \, \big] - k \sqrt{-G} \,\, \big[\, (\partial_A \phi )(\partial_B \phi ) - \frac{1}{2} G_{AB} (\nabla \phi)^2 \,\big]~~~~~~~~~ \nonumber \\
+\,\frac{1}{2} \sqrt{-G}\, G_{AB} \, e^{b\phi} \lambda + \sum_{i=1}^{2} \kappa^2 \sqrt{-g}\, G_{\mu A} G^{\mu\nu}G_{\nu B} T^{(i)} (\phi) \delta (y-y_i ) =0 \,\,,
\end{eqnarray}
and the equation for $\phi$
\begin{equation}
\sqrt{-G} \,\,\big[\, 2k \,\,\Box \phi - b \, e^{b\phi} \lambda \,\big] - \sum_{i=1}^{2} 2\kappa^2 \sqrt{-g}\, \,\frac{\partial T^{(i)}(\phi)}{\partial \phi} \,\delta (y-y_i ) =0 \,\,,
\end{equation}
where $A,B=0,1,2,3,5$, while $\mu,\nu=0,1,2,3$. Let us introduce the 5d metric in the form
\begin{equation}
 ds^2_{\rm E} = \frac{dy^2}{f^2 (y)} + e^{2 A(y)} \big [\,-dt^2 + d {\vec{x}_3}^{\,2} \,\big]\,\,.
\end{equation}
In (2.6), $f(y)$ is an extra degree of freedom associated with the coordinate transformation $y \rightarrow y^{\prime}=F(y)$, so it can be taken arbitrarily as we wish. The complete set of field equations are then given by
\begin{equation}
 3A^{\prime\prime} + 6 {A^{\prime}}^2 + 3 A^{\prime} \frac{f^{\prime}}{f} + \frac{k}{2}{\phi^{\prime}}^2 +\frac{1}{2} \frac{e^{b\phi}}{f^2}\lambda = - \sum_{i=1}^{2} \kappa^2 \frac{1}{f} T^{(i)}(\phi) \,\delta (y-y_i )  \,\,, \end{equation}
\begin{equation}
6 {A^{\prime}}^2 - \frac{k}{2}{\phi^{\prime}}^2 + \frac{1}{2} \frac{e^{b\phi}}{f^2}\lambda = \sum_{i=1}^{2} 2\kappa^2 \frac{1}{f} \frac{\partial T^{(i)}(\phi)}{\partial \phi} \,\delta (y-y_i )  \,\,,
\end{equation}
\begin{equation}
2k\,\big(\phi^{\prime\prime} + 4 A^{\prime} \phi^{\prime} + \frac{f^{\prime}}{f} \phi^{\prime}\,\big)-b \, \frac{e^{b\phi}}{f^2}\lambda = \sum_{i=1}^{2} 2\kappa^2 \frac{1}{f} \frac{\partial T^{(i)}(\phi)}{\partial \phi} \,\delta (y-y_i )  \,\,,
\end{equation}
where the "prime" denotes the derivative with respective to $y$. In the above equations the first two are the ($\mu\nu$) and ($55$) components of the Einstein equation, while the last one follows from the equation for $\phi$.

Taking linear combination of (2.7) and (2.8), and choosing $f^2 \,=\, e^{b\phi}$ one finds that the above set of field equations reduces to
\begin{equation}
\frac{d}{dy} \,\Big( \, \xi \frac{dA}{dy} \,\Big) + \frac{1}{3} \lambda \xi = - \sum_{i=1}^{2} \frac{\kappa^2}{3} C^{(i)}_1  \,\delta (y-y_i ) \,\,,
\end{equation}
\begin{equation}
\frac{d}{dy} \,\Big(\, \xi \frac{d\phi}{dy} \,\Big) - \frac{b}{2k} \lambda \xi = \sum_{i=1}^{2} \frac{\kappa^2}{k} C^{(i)}_2  \,\delta (y-y_i ) \,\,,
\end{equation}
\begin{equation}
6 \Big(\frac{dA}{dy} \Big)^2 - \frac{k}{2} \Big(\frac{d\phi}{dy} \Big)^2 + \frac{1}{2} \lambda =0 \,\,,
\end{equation}
where $\xi$ is defined by
\begin{equation}
\xi \,=\, e^{4A} f \,=\, e^{4A +\frac{b}{2}\phi} \,\,,
\end{equation}
while $C^{(i)}_k$ ($k=1,2$) are
\begin{equation}
C^{(i)}_1 = e^{4A} \,\, T^{(i)} (\phi)\Big|_{y=y_i} \,\,,~~~~~C^{(i)}_2 = e^{4A} \,\, \frac{\partial T^{(i)}(\phi)}{\partial \phi}\Big|_{y=y_i} \,\,.
\end{equation}

\vskip 1cm
\hspace{-0.7cm} {\bf 3. Solution to field equations with $\lambda \neq 0$}
\renewcommand{\theequation}{3.\arabic{equation}}
\setcounter{equation}{0}
\vskip 0.5cm
\hspace{-0.6cm}By inspecting (2.13) together with (2.10) and (2.11) one finds that $\xi$ must satisfy
\begin{equation}
\frac{d^2 \xi}{d y^2} + \beta \lambda \xi = \sum_{i=1}^{2} a^{(i)}_{\xi} \, \delta(y-y_i)\,\,,
\end{equation}
where
\begin{equation}
\beta = \frac{4}{3} - \frac{b^2}{4k}\,\,,
\end{equation}
and
\begin{equation}
a^{(i)}_{\xi}= -\frac{\kappa^2}{2}\,\frac{b}{k} \Big[\, \frac{8}{3}\frac{k}{b} C^{(i)}_1 -  C^{(i)}_2 \Big] \,\,.
\end{equation}
The solution to the differential equations (2.10) and (2.11) can be readily found. Using (3.1), one can show that the most general solution to the set of field equations (2.10) and (2.11) takes the form
\begin{equation}
e^{A} = i_A \, \xi^{\frac{1}{3\beta}}\,\,,~~~~~~e^{\phi} =  i_\phi \, \xi^{-\frac{b}{2k\beta}} \,\,,
\end{equation}
where $i_M$ ($M=A, \phi$) are defined by the equation
\begin{equation}
\frac{d}{dy}\Big( \frac{\xi}{i_M} \frac{di_M}{dy} \Big) = \sum_{i=1}^{2} a^{(i)}_{M} \, \delta (y-y_i )
\end{equation}
with $a^{(i)}_{M}$ given by
\begin{equation}
a^{(i)}_{A} =  \frac{b}{6}\kappa^2 \frac{1}{k\beta} \Big[ \frac{b}{2}\, C^{(i)}_1 - C^{(i)}_2 \Big] \,\,,~~~a^{(i)}_{\phi} = - \frac{4}{3}\kappa^2 \frac{1}{k\beta} \Big[ \frac{b}{2} \, C^{(i)}_1 - C^{(i)}_2 \Big] \,\,.
\end{equation}

Now we want to find $\xi$ and $i_M$ satisfying the boundary condition at $y=y_1$ (i.e., at $y=0$). First we consider the case $\lambda \neq 0$. (Throughout this paper we mainly consider the case $\lambda \neq 0$. The case $\lambda =0$ is separately discussed in Sec.6.) With an assumption of the $\mathbb{Z}_2$-symmetry $y \rightarrow -y$, one can show that the most general solution to (3.1) valid in the region $-\pi R_c <y<\pi R_c$ appears to be
\begin{equation}
\xi = \xi_0 \, e^{-\sqrt{- \beta  \lambda} \,|y|} + \xi_1 \cosh {\sqrt{- \beta \lambda} \, y} \,\,,
\end{equation}
where $\xi_1$ is an arbitrary constant, but $\xi_0$ is given by
\begin{equation}
\xi_0 = - \frac{a^{(1)}_\xi}{2 \sqrt{- \beta \lambda}} \,\,\,.
\end{equation}
Also the $\mathbb{Z}_2$-symmetry requires that $i_M$ defined by (3.5) must satisfy
\begin{equation}
\frac{\xi}{i_M} \frac{di_M}{dy} = \frac{a^{(1)}_M}{2} \,\epsilon (y)
\end{equation}
where $\epsilon (y)$ is the step function defined by
\begin{equation}
\epsilon (y) = \left \{ \begin{array}{l}
                   +1\,\,,~~~~{\rm for}~~y>0 \\
                   -1\,\,,~~~~{\rm for}~~y<0 \,\,.
               \end{array} \right.
\end{equation}

We have seen that (3.4) is the solution to (2.10) and (2.11). But we still need for consistency to check whether it satisfies (2.12) either before we proceed further. Substituting (3.4) (together with (3.7) and (3.9)) into (2.12) gives two consistency conditions
\begin{equation}
8\,a^{(1)}_A + b\, a^{(1)}_{\phi} = 0 \,\,,
\end{equation}
and
\begin{equation}
3\,{a^{(1)}_A }^2 - \frac{k}{4}\, {a^{(1)}_{\phi} }^2 + \lambda \,(\xi^2_1 + 2 \xi_0 \xi_1 )=0 \,\,.
\end{equation}
Among these two conditions the first one is identically satisfied by (3.6). But (3.12) (upon using (3.8) and (3.11)) gives a condition
\begin{equation}
\xi_1 = \xi_0 \bigg[-1 \pm \,\sqrt{1- \frac{3}{4}\, k \beta^2 l^2}\,\,\, \bigg] \,\,,
\end{equation}
where $l$ is defined by
\begin{equation}
l = {a^{(1)}_{\phi}}/{a^{(1)}_{\xi}} \,\,.
\end{equation}
(3.13) contains two roots for $\xi_1$. But only the upper sign agrees with the RS-model (upon taking $k=0$) as will be clear later (see (4.2)). Thus in what follows we will only consider the case where $\xi_1$ is given by $\xi_1 =\xi_0 \big[-1 +\sqrt{1- \frac{3}{4}k \beta^2 l^2}\,\,\big]$.

To find $i_M$, it is convenient to rewrite $\xi$ as
\begin{equation}
\xi = \left \{ \begin{array}{l}
                   c_1 \, e^{\sqrt{- \beta  \lambda} \,y} + c_2 \, e^{-\sqrt{- \beta  \lambda} \,y} ~~~~(y>0) \\ \\
                   c_2 \, e^{\sqrt{- \beta  \lambda} \,y} + c_1 \, e^{-\sqrt{- \beta  \lambda} \,y}~~~~(y<0)
               \end{array} \right.
\end{equation}
with $c_1 = \frac{\xi_1}{2}$ and $c_2 = \xi_0 + \frac{\xi_1}{2}$, respectively. Substituting (3.15) into (3.9) one finds
\begin{equation}
i_M = d_M \Bigg( \frac{1 + \sqrt{p}\, e^{\sqrt{- \beta  \lambda} \,|y|}}{1 - \sqrt{p}\,e^{ \sqrt{- \beta  \lambda} \,|y|}} \Bigg) ^{I_M} \,\,\,,
\end{equation}
where $d_M$ is an arbitrary constant, while
\begin{equation}
p= \frac{1- \sqrt{1-\frac{3}{4}k\beta^2 l^2}}{1+ \sqrt{1-\frac{3}{4}k\beta^2 l^2}} \,\,,
\end{equation}
and
\begin{equation}
I_M = \,- \,\big(\,\frac{3}{4}k \,\big)^{-\frac{1}{2}}\,\frac{\epsilon_l}{|\beta|}\, \frac{a^{(1)}_{M}}{a^{(1)}_{\phi}} \,\,,
\end{equation}
where $\epsilon_l \equiv l/|l|$.

\vskip 1cm
\hspace{-0.7cm} {\bf 4. Warp factor and 5d geometry}
\renewcommand{\theequation}{4.\arabic{equation}}
\setcounter{equation}{0}
\vskip 0.5cm
\hspace{-0.6cm}In the framework of string theory the 5d metric in (2.6) corresponds to the Einstein metric and it is related to the string metric by the equation $ds^2_{\rm E}=e^{-b\phi}\, ds^2_{\rm string}$. So the string metric takes the form
\begin{equation}
ds^2_{\rm string} = dy^2 + W(y) [-dt^2 + d\vec{x}^2_3 \,]
\end{equation}
with the warp factor $W(y)$ given by
\begin{eqnarray}
W(y) = d^2_A \, d^b_{\phi} \, \Bigg( \frac{1 + \sqrt{p}\, e^{\sqrt{- \beta  \lambda} \,|y|}}{1 - \sqrt{p}\,e^{ \sqrt{- \beta  \lambda} \,|y|}} \Bigg) ^{I_W} \Big[ \xi_0 e^{-\sqrt{- \beta  \lambda} \,|y|} + \xi_1 \cosh \sqrt{- \beta  \lambda} \,y \Big]^{2(1-\frac{1}{\beta})} \,\,,
\end{eqnarray}
where $I_W = 2I_A + bI_{\phi}$, and therefore by (3.11) and (3.18)
\begin{equation}
I_W = -\, \big(\frac{3}{4}b\big) \, \big(\frac{3}{4}k \big)^{-\frac{1}{2}} \,\frac{\epsilon_l}{|\beta|} \,\,.
\end{equation}
So in the string theory limit ($b=k=\frac{4}{3}$, $\beta=1$) $W(y)$ is given by
\begin{equation}
W(y)=d^2_A \, d^{\frac{4}{3}}_\phi \, \Bigg( \frac{1+\sqrt{p} e^{\sqrt{-\lambda}|y|}}{1-\sqrt{p} e^{\sqrt{-\lambda}|y|}} \Bigg)^{-\epsilon_l} \,\,\,.
\end{equation}
Note that the terms in $\xi$ do not contribute to $W(y)$ in this case.

In the case of the RS-model ($b=k^{1/2}=0$, $\beta=\frac{4}{3}$), however, there is no distinction between $ds^2_{\rm E}$ and $ds^2_{\rm string}$, and the metric is simply represented by (4.1) together with (4.2). Indeed (4.2) becomes (upon taking $d_A = \xi^{-1/4}_0$)
\begin{equation}
W(y) = e^{-\sqrt{-\frac{1}{3}\lambda}\,|y|}
\end{equation}
for $b=k^{1/2}=0$, which coincides with the solution in \cite{17}.

\vskip 1cm

\hspace{-0.7cm} {\bf 5. Boundary condition at $y= \pi R_c$}
\renewcommand{\theequation}{5.\arabic{equation}}
\setcounter{equation}{0}
\vskip 0.5cm
\hspace{-0.6cm}In Sec.3, we have considered the boundary condition at $y=0$. Now we require that the solution also satisfy the boundary condition at $y=\pi R_c$ either. To consider the boundary condition at $y= \pi R_c$ we need to know the singular behaviors of $\xi (y)$ and $i_M (y)$ at that point. Note that in order for $\xi(y)$ to be a periodic with a period $2\pi R_c$, $\xi(y)$ must be given in the region $0<y<2\pi R_c$ by
\begin{equation}
\xi = \left \{ \begin{array}{l}
                   \xi_0 \, e^{-\sqrt{- \beta  \lambda} \,y} + \xi_1 \, \cosh {\sqrt{- \beta  \lambda} \,y} ~~~~(0<y<\pi R_c) \\ \\
                   \xi_0 \, e^{\sqrt{-\beta  \lambda} \,(y-2\pi R_c )} + \xi_1 \, \cosh {\sqrt{- \beta  \lambda} \,(y-2\pi R_c )}~~~~(\pi R_c <y<2\pi R_c ) \,\,,
               \end{array} \right.
\end{equation}
and in terms of a new coordinate $\tilde{y} \equiv y- \pi R_c$ (5.1) can be rewritten as
\begin{equation}
\xi (\tilde{y}) = \xi_0 \, e^{-\sqrt{- \beta  \lambda} \,(\pi R_c -|\tilde{y}|)} + \xi_1 \, \cosh {\sqrt{- \beta  \lambda} \,(\pi R_c - |\tilde{y}|)} \,\,,
\end{equation}
which shows that $\xi (\tilde{y})$ is singular at the point $\tilde{y}=0$ ($y= \pi R_c$).
Besides this, in the region $0<y<2\pi R_c$ (3.1) also reduces to
\begin{equation}
\frac{d^2 \xi}{d \tilde{y}^2} + \beta \lambda \xi = a^{(2)}_\xi \delta (\tilde{y}) \,\,.
\end{equation}
Substituting (5.2) into (5.3), and using (3.8) and (3.13) one finds a condition
\begin{equation}
a^{(2)}_\xi = - \, a^{(1)}_\xi e^{-\sqrt{- \beta  \lambda} \,\pi R_c} \Sigma (l) \,\,,
\end{equation}
where
\begin{equation}
\Sigma(l) = 1+ \Big(1- \sqrt{1-\frac{3}{4} k \beta^2 l^2 } \,\Big) \,e^{\sqrt{- \beta  \lambda} \,\pi R_c} \sinh{\sqrt{- \beta  \lambda} \,\pi R_c}  \,\,.
\end{equation}

Similarly, in order for $i_M$ to be periodic with a period $2\pi R_c$ it must be of the form
\begin{equation}
i_M (\tilde{y}) = d_M \Bigg ( \frac{1+\sqrt{p}\,e^{\sqrt{- \beta  \lambda} \,(\pi R_c -|\tilde{y}|)}}{1-\sqrt{p}\,e^{\sqrt{- \beta  \lambda} \,(\pi R_c -|\tilde{y}|)}} \Bigg)^{I_M}
\end{equation}
in the region $0<y<2\pi R_c$. Substituting (5.6) into the equation
\begin{equation}
\frac{\xi}{i_M} \frac{di_M}{d\tilde{y}} = \frac{a^{(2)}_M}{2} \epsilon (\tilde{y}) \,\,,
\end{equation}
we find the second condition
\begin{equation}
a^{(2)}_M = - a^{(1)}_M \,\,.
\end{equation}

Now we assume that $T^{(i)}(\phi)$ is given by (2.3). Then since $C^{(i)}_2 = \alpha \, C^{(i)}_1$ for such $T^{(i)}(\phi)$, (5.4) and (5.8) reduce, respectively, to
\begin{equation}
\Big(\, \frac{8k}{3b}-\alpha \Big)\, C^{(2)}_1 = -\Big(\,\frac{8k}{3b}-\alpha \Big)\, C^{(1)}_1 e^{-\sqrt{- \beta  \lambda} \,\pi R_c} \, \Sigma (l) \,\,,
\end{equation}
and
\begin{equation}
\Big(\, \frac{b}{2}-\alpha \Big)\, C^{(2)}_1 = -\Big(\,\frac{b}{2}-\alpha \Big)\, C^{(1)}_1 \,\,.
\end{equation}
In general (5.9) and (5.10) are incompatible\footnote{To be precise, both (5.9) and (5.10) can be satisfied simultaneously for an arbitrary $\alpha$ if $e^{-\sqrt{-\beta\lambda}\pi R_c} \Sigma(l)=1$. But this equation is only solved by $e^{\sqrt{-\beta\lambda}\pi R_c}=1$ or $e^{\sqrt{-\beta\lambda}\pi R_c} = [1+\sqrt{1-\frac{3}{4}k \beta^2 l^2}\,]/ [1-\sqrt{1-\frac{3}{4}k \beta^2 l^2}\,]$, where the first equation implies $R_c =0$, while the second equation yields $R_c =\infty$ for both $k=0$ (RS case) and $l=0$ (NS-brane). Finally for the D-brane ($k=\frac{4}{3}$, $\beta=1$, $l=-2$) the second equation does not yield a real solution for $R_c$. For these reasons, we discard the case where $\alpha$ is arbitrary.} with one another unless $\alpha = {8k}/{3b}$ or $\alpha = {b}/{2}$. So $\alpha$ must be one of these values. But if $\alpha = {8k}/{3b}$, $l \rightarrow \infty$ from (3.3) and (3.14). So we discard it and we only take
\begin{equation}
\alpha = \frac{b}{2} \,\,.
\end{equation}
By (5.11), (5.10) becomes an empty condition and we are only left with (5.9) which now reads
\begin{equation}
C^{(2)}_1 = - C^{(1)}_1 e^{-\sqrt{-\beta\lambda}\pi R_c}
\end{equation}
because $l$ vanishes for $\alpha=b/2$. Further, using (2.13) and the fact that $\xi$ becomes $\xi =\xi_0 e^{-\sqrt{-\beta\lambda}|y|}$ for $l=0$, one finds from (5.12) that
\begin{equation}
T^{(2)}_0 = - T^{(1)}_0 \,\,.
\end{equation}
That is, the tensions of the branes located at the orbifold fixed points $y=0$ and $\pi R_c$ must be equal in magnitude but opposite in sign from one another.

\vskip 1cm
\hspace{-0.7cm} {\bf 6. $\lambda = 0$ case}
\renewcommand{\theequation}{6.\arabic{equation}}
\setcounter{equation}{0}
\vskip 0.5cm
\hspace{-0.6cm}So far we have only considered the case $\lambda \neq 0$. In this section we want to find the solution to the field equations with $\lambda =0$. In the case $\lambda =0$, the field equations are the same as before except $\lambda$ be set equal to zero everywhere, and the solution to the field equations is now given by
\begin{equation}
e^A =i_A, ~~~e^{\phi}=i_{\phi} \,\,,
\end{equation}
where $i_M$ are still defined by (3.5) (and therefore by (3.9) and (5.7) for the $\mathbb{Z}_2$-symmetric configurations) except that $a_M^{(i)}$ are replaced by $\tilde{a}_M^{(i)}$:
\begin{equation}
\tilde{a}_A^{(i)} = -\frac{\kappa^2}{3}C_1^{(i)}, ~~~~~\tilde{a}_{\phi}^{(i)} = \frac{\kappa^2}{k}C_2^{(i)} \,\,,
\end{equation}
and $\xi$ is now the solution of
\begin{equation}
\frac{d^2 \xi}{dy^2} = a_{\xi}^{(1)} \delta (y) \,\,.
\end{equation}
The general solution to (6.3) satisfying both the $\mathbb{Z}_2$-symmetry and boundary condition at $y=0$ is simply
\begin{equation}
\xi (y) = \tilde{\xi}_0 |y| + \tilde{\xi}_1 \,\,,
\end{equation}
where $\tilde{\xi}_1$ is an arbitrary constant, but $\tilde{\xi}_0$ is related to $a_{\xi}^{(1)}$ by the equation
\begin{equation}
\tilde{\xi}_0 = \frac{a_{\xi}^{(1)}}{2} \,\,.
\end{equation}
So from (3.9) one obtains
\begin{equation}
i_M =d_M (\tilde{\xi}_0 |y| + \tilde{\xi}_1 )^{\tilde{l}_M} \,\,,
\end{equation}
where $\tilde{l}_M$ are defined by $\tilde{l}_M = \tilde{a}_{M}^{(1)} / a_{\xi}^{(1)}$.

Now we have to check that this solution also satisfies the consistency condition (2.12) with $\lambda = 0$. Substituting the above solution into (2.12), one obtains a condition
\begin{equation}
6 {\tilde{a}_{A}^{(1)}}{ }^2 - \frac{k}{2} {\tilde{a}_{\phi}^{(1)}}{ }^2=0  \,\,,
\end{equation}
which, upon setting $T(\phi)=T_0 e^{\alpha \phi}$, reduces to
\begin{equation}
\alpha^2 = \frac{4}{3} k\,\,.
\end{equation}
In the string theoretical set up this result is obscure. For $k=\frac{4}{3}$, (6.8) gives $\alpha=\pm \frac{4}{3}$ which corresponds to neither the NS-brane nor D-brane of the string theory. (6.8) implies that the branes on the orbifold fixed points must appear in the form of a very specific combination of the NS- and D-branes. Indeed, the natural interpretation for this value of $\alpha$ may be that the scalar field $\phi$ would not be identified as a dilaton of the string theory in the case of $\lambda =0$. The lack of natural string theoretical interpretation in the case $\lambda =0$ may be due to the fact that the action (2.1) does not have enough terms characterizing the string theory in the absence of $\lambda$ term with the factor $e^{\frac{4}{3}\phi}$. Since the natural string theoretical interpretation does not exist in the case $\lambda =0$, in the following discussion we will only consider the case $\lambda \neq 0$.

In the case $\lambda =0$, the boundary conditions at $y=\pi R_c$ require
\begin{equation}
a_\xi^{(2)}= -a_\xi^{(1)}\,\,,~~~~~~~~~~\tilde{a}_{M}^{(2)} = - \tilde{a}_{M}^{(1)}
\end{equation}
which are satisfied only if
\begin{equation}
C_{1}^{(2)} = - C_{1}^{(1)} \,\,.
\end{equation}

\vskip 1cm
\hspace{-0.7cm} {\bf 7. Background branes at orbifold fixed points}
\renewcommand{\theequation}{7.\arabic{equation}}
\setcounter{equation}{0}
\vskip 0.5cm
\hspace{-0.6cm}Turning back to Sec.5 it should be noted that the condition (5.11) is very important because it tells us of what type the brane under consideration would be. For instance $b=0$ gives $\alpha=0$ which of course corresponds to the brane of the RS-model. In the string theoretical setup, on the other hand, (5.11) gives $\alpha=\frac{2}{3}$ indicating that the branes placed at the orbifold fixed points must be basically NS-branes.\footnote{A similar example of such a configuration can be found in \cite{27-1} where the author observed that the world-sheet CFT of the type IIA orbifold $R_6 \times R_4 / {I}_2 $ describes a solitonic fivebrane (NS5-brane) living in the fixed point of $R_4 / \mathbb{Z}_2$.} Moreover, as mentioned in \cite{22} the dynamics of the world volume degrees of freedom on the NS-brane does not depend on the dilaton, meaning that the relevant coupling constant is dilaton independent, and consequently $T^{(i)}(\phi)$ is expected to maintain the tree level form even under quantum corrections to the brane tension due to dynamics of world volume fields. Thus the above argument is not restricted only to the tree level string theory.

The configuration where NS-branes reside at the orbifold fixed points has interesting aspects. For the NS-branes $T^{(i)}(\phi)$ is given by (2.3) with $\alpha=\frac{b}{2}=\frac{2}{3}$, so $a^{(i)}_{\phi}$ and consequently $l$, $p$, and $\xi_1$ all vanish ; $l=p=\xi_1=0$. Also since $\beta=1$ for $b=k=\frac{4}{3}$, one finds that $W(y)$ in (4.2) is simply a constant, i.e.,  $W(y)=1$ if we choose $d^2_A d^b_{\phi} =1$, and therefore (4.1) becomes a flat metric, a direct product of $M_4$ and $S_1 /\mathbb{Z}_2$, where $M_4$ is the 4d Minkowski space. This is interesting. In the string frame the 5d spacetime does not know about the existence of the constant $\lambda$. In the presence of the background NS-branes it remains flat even for nonzero $\lambda$. The 5d metric only acquires a conformal factor $e^{-b \phi}$ in the Einstein frame, where the string coupling $e^{\phi}$ is given (for $b=k=\frac{4}{3}$ and $l=0$) by\footnote{The constant $g_s$ was defined by $g_s = e^{\phi} \big|_{y=0}$, which amounts to choosing $d_{\phi} =g_s \xi^{\frac{1}{2}}_0$.}
\begin{equation}
e^{\phi}= g_s \, e^{\sqrt{-\lambda} |y|/2} \,\,.
\end{equation}

The above discussion leads to an important result. The flat metric $M_4 \times S_1/\mathbb{Z}_2$ can be obtained only through an introduction of the background NS-branes on the orbifold fixed points. Indeed, without these background branes the 5d metric becomes singular everywhere as can be readily checked as follows. In the absence of branes $a_\xi$ and $a_M$ all vanish, and from (3.8) and (3.9) one finds that $\xi_0 =0$ and $i_M = \rm constant$. Also since $\xi_1 =0$ (for $\lambda \neq 0$ and $a_M =0$) from (3.12), $\xi$ and consequently the conformal factor $e^{-b\phi}$ vanish everywhere in $S_1$. This property is very reminiscent of the $(p+3)$d brane world models \cite{1} in which the presence of NS-NS type $p$-brane is indispensable to obtain background $R_2$ or $R_2/\mathbb{Z}_n$ on the transverse dimensions. Without this brane the 2d transverse space becomes a "semi-infinitely long pin" whose tip is identified as the origin of the coordinates. The NS-NS type $p$-brane (with negative tension) is necessary to spread out this pin-shaped space to get $R_2$ or $R_2/\mathbb{Z}_n$. In general codimension-1 brane solutions take completely different forms as compared with codimension-2 brane solutions, and the above correspondence between 5d models and $(p+3)$d models suggests that the role of the NS-branes in the background geometries may be much more significant than we expect.

\vskip 1cm
\hspace{-0.7cm} {\bf 8. Towards the cosmological constant problem}
\renewcommand{\theequation}{8.\arabic{equation}}
\setcounter{equation}{0}
\vskip 0.5cm
\hspace{-0.6cm}The consequence of the fact that the spacetime contains background NS-branes is crucial in the context of the cosmological constant problem. Consider a configuration that a D-brane is introduced on the background NS-brane, and SM-fields are living on that D-brane. In string theory the tension of the codimension-1 D-brane is expected to be given \cite{22} by a power series of the form $T_D (\phi) = e^{(5/3)\phi} \sum_{n=0}^{\infty} T^{(D)}_n e^{n \phi}$, and therefore the tension of the coincident brane (of the background NS-brane and the D-brane) would roughly take the form\footnote{To be precise, the tension of the bound state of the D-brane and NS-brane may appear rather complicated than this if their world volumes are in parallel with each other. It is known \cite{28} that in this case the supersymmetry of the system is completely broken and the D-brane becomes unstable and eventually decays losing most of its energy. However, if some of dimensions of their world volumes are not in parallel then the decay of the D-brane can be avoided. Also see \cite{29}.}
\begin{equation}
T(\phi) = e^{\frac{2}{3}\phi} \, T^{(NS)}_0 + e^{\frac{5}{3}\phi}\, T^{(D)}_0 + \sum_{n=1}^{\infty} e^{(\frac{5}{3}+n) \phi}\, T^{(D)}_n  \,\,,
\end{equation}
where $T^{(NS)}_0$ and $T^{(D)}_0$ are both of an order $\sim 1/{\alpha^{\prime}}^2$, and the terms with $n \geq 1$ represent the quantum correction terms. Substituting (8.1) into (3.14) one finds
\begin{equation}
l= - g_s \, \frac{\Big[T^{(D)}_0 + \sum_{n=1}^{\infty} (n+1)T^{(D)}_n g^n_s \Big]}{\Big[ T^{(NS)}_0 +\frac{1}{2} T^{(D)}_0 g_s - \sum_{n=1}^{\infty} \frac{(n-1)}{2} T^{(D)}_n g^{(n+1)}_s   \Big]} \,\,,
\end{equation}
where $g_s$ is defined by $g_s \equiv e^{\phi} \big|_{y=0}$ (see footnote 3).

With the background NS-brane alone (8.2) gives $l=0$ as we already know. Now introduce a D-brane (SM-brane) on the background NS-brane. In this case $l$ acquires a nonzero values. Neglecting the quantum correction terms one finds
\begin{equation}
l \sim - \Big( T^{(D)}_0 / T^{(NS)}_0 \Big) g_s \,\,.
\end{equation}
(8.3) implies that $l$ is as small as $g_s$, $l \sim g_s$, since $T^{(D)}_0$ and $T^{(NS)}_0$ are of the same order $\sim 1/{\alpha^{\prime}}^2$. Thus in the limit $g_s \rightarrow 0$ the effect of introducing D-brane on the bulk geometry is negligibly small. Indeed, (4.2) can be expanded in a power series of $l$ as
\begin{eqnarray}
W(y)=d^2_A \, d^b_{\phi} \xi^{2(1-\frac{1}{\beta})}_0 \, e^{-2(1-\frac{1}{\beta}) \sqrt{-\beta \lambda}|y|}\,\Big[1-\big(\,\frac{3}{4}b\big)\,l\,e^{\sqrt{-\beta \lambda}|y|} + \frac{1}{2}\big(\,\frac{3}{4}b\big)^2\,l^2\,e^{2\sqrt{-\beta \lambda}|y|} \nonumber \\
- \beta^2 (1-\frac{1}{\beta}) \big(\,\frac{3}{4}k\big)\,l^2 \, e^{\sqrt{-\beta \lambda}|y|}\,\cosh{\sqrt{-\beta \lambda}|y|} + O(l^3)\, \Big]\,\,. ~~~~
\end{eqnarray}
In the RS limit ($b=k^{1/2}=0$, $\beta=\frac{4}{3}$) (8.4) simply becomes (4.5) as it should be. In this case $W(y)$ is not given by a power series of $l$. In the string theory limit ($b=k=4/3$, $\beta=1$), however, (8.4) becomes
\begin{equation}
W(y) = 1-l\,e^{\sqrt{-\lambda}|y|}\,+\frac{1}{2}l^2 \, e^{2\sqrt{-\lambda}|y|}\,  + O(l^3) \,\,
\end{equation}
upon taking $d^2_A d^{4/3}_\phi =1$, which shows that the change of $W(y)$ due to an introduction of the D-brane is only of an order $\sim g_s$.

Once $l$ is determined at the tree level, the effect of the higher order terms can be obtained by adding $\delta l$ to $l$, where $\delta l$ is the shift in $l$ due to the quantum correction (the terms with $n \geq 1$) to the D-brane tension. From (8.2) one finds that $\delta l$ is only of an order $\sim g^2_s$ :
\begin{equation}
 \delta l \sim \Big( T^{(D)}_1 / T^{(NS)}_0 \Big) g^2_s  \,\,,
\end{equation}
which implies that the change of the bulk geometry due to quantum fluctuations of SM-fields with support on the D-brane is extremely suppressed in the limit $g_s \rightarrow 0$. In the brane world models the intrinsic curvature of the brane is $\it{a~priori}$ zero, and consequently the whole quantum fluctuations of SM-fields entirely contribute to changing the geometry of bulk space. Thus in general the bulk geometry is necessarily severely disturbed by the quantum fluctuations. In our case, however, the disturbance due to quantum fluctuations is highly suppressed as mentioned above, and the bulk geometry, as well as the flat geometry of the brane, is virtually insensitive to the quantum fluctuations. Such a feature also can be found in the ($p+3$)d brane world models \cite{9}, and provides a new type of self-tuning mechanism with which to solve the cosmological constant problem. See Sec.9.

\vskip 1cm
\hspace{-0.7cm} {\bf 9. Summary and discussion}
\renewcommand{\theequation}{9.\arabic{equation}}
\setcounter{equation}{0}
\vskip 0.5cm
\hspace{-0.6cm}In this paper we have studied codimension-1 brane solutions of the 5d brane world models compactified on $S_1 / \mathbb{Z}_2$. In string theoretical setup they suggest that background branes sitting at orbifold fixed points should be NS-branes, rather than D-branes. Indeed, the existence of the background NS-branes is indispensable to obtain the flat geometry $M_4 \times S_1 / \mathbb{Z}_2$, and without these branes the 5d metric becomes singular (it vanishes everywhere in the Einstein frame), indicating that the flat $M_4 \times S_1 / \mathbb{Z}_2$ inherently involves the NS-branes in itself. The ansatz for the 5d metric of this paper is the most general ansatz we can think of for the configurations of 5d spacetime with Poincar$\acute{\rm e}$ symmetry in the 4d subsector, which suggests that the solutions obtained in this paper would be the one that is closer to the true extremum of the action than the others. In general fixing an ansatz leads to a limited class of geometries.

The same result also can be found in the $(p+3)$d effective string theory where the $p$-brane appears as a codimension-2 brane. In \cite{1} it was argued that in $(p+3)$d string theory the existence of NS-NS type $p$-brane is indispensable to obtain background geometry $R_2$ or $R_2 /\mathbb{Z}_n$ on the transverse dimensions. In the absence of this background brane the 2d transverse space becomes a "semi-infinitely long pin", and the only way to avoid this singular space is to introduce NS-NS type $p$-brane with negative tension on the orbifold fixed point. In general codimension-1 brane theory is entirely different from the codimension-2 brane theory, and the above correspondence between these two entirely different theories leads us to a conjecture that the whole flat backgrounds of the string-inspired brane world models (or the string theory itself) inherently involve the NS-branes implicitly in their ansatz.

As mentioned above, in the presence of the NS-branes the 5d spacetime becomes a direct product $M_4 \times S_1 / \mathbb{Z}_2$ with  $\mathbb{Z}_2$-symmetry in the fifth compact dimension. The NS-branes contribute to fixing the background geometry of the 5d spacetime. After the geometry of the spacetime is fixed by the NS-branes we may also introduce D-branes on this background $M_4 \times S_1 / \mathbb{Z}_2$. Introducing D-branes, however, hardly affects the background geometry. It affects the 5d geometry only to the extent of an order $\sim g_s$, and similarly the quantum fluctuations of SM-fields to the extent of an order $\sim g^2_s$, which are entirely negligible in the limit $g_s \rightarrow 0$.

In the brane world models the intrinsic curvature of the brane is $\it {a~priori}$ zero, and the 4d cosmological constant automatically vanishes, leading to a natural solution to the cosmological constant problem. But in this case the inherently fixed geometry of the brane gives rise to an unwanted problem. The whole quantum fluctuations of SM-fields entirely contribute to changing the geometry of bulk space because the geometry of the brane is already fixed to be flat from the beginning. Thus the bulk geometry, and therefore the compactification scale $R_s$ in the 5d case is necessarily severely disturbed by the quantum fluctuations, and this disturbance of the compactification scale can in turn lead to severe disturbances of the observed coupling constants. In the given configurations of this paper such a problem can be naturally solved. In the presence of the background NS-branes the disturbances due to quantum fluctuations is highly suppressed in the limit $g_s \rightarrow 0$, and consequently the coupling constants are not disturbed by the quantum fluctuations of the SM-fields living on the D-branes. Such a feature may lead to a new type of brane world model with which to solve the cosmological constant problem.

In our configurations the natural string theoretical interpretation does not exist when $\lambda =0$. So we only consider the case $\lambda \neq 0$, and in this case the magnitude of $\lambda$ can be readily estimated in the string theoretical setup. Namely, $l$ and therefore $\xi_1$ vanish for $\alpha ={b}/{2}$, so using (2.13) and (3.3) one finds from (3.8) that $\sqrt{-\lambda}= \kappa^2 T^{(1)}_{0}/2$ for $k=b=\frac{4}{3}$ and $\beta =1$, or since $\kappa^2 \sim 1/M^3_5$ and $T^{(1)}_{0} \sim 1/{\alpha^{\prime}}^2 \sim M^4_s$, where $M_5$ and $M_s$ are the 5d Planck scale and the string scale respectively, $\sqrt{-\lambda}$ can be rewritten as   $\sqrt{-\lambda} \sim {\kappa^2}/{{\alpha^{\prime}}^2} \sim M^4_s /M^3_5$. In string theory $\lambda$ is only of a subleading order and vanishes at the tree level in ordinary circumstances. But still one can consider more general cases where $\lambda$ can have nonzero values at the tree level. For instance closed string backgrounds with nonzero tree-level cosmological constant can be obtained by considering subcritical strings which could arise naturally as a result of tachyon condensation \cite{30,31}. But in this case $\lambda$ is given by $\lambda \sim - 1/{\alpha}^{\prime}$, and in order for this to be consistent with the above $\lambda$, $M_5$ must be proportional to $M_s$, i.e., $M_5 = \gamma M_s$ with $\gamma \sim O(1)$.\footnote{This result coincides with the ($p+3$)dimensional case of \cite{1} where $M_s$ and $M_{p+3}$ must be of the same order, $M_s \sim M_{p+3}$ in order that the background geometry of the 2d transverse space becomes $R_2$ or $R_2 / \mathbb{Z}_n$.}

Apart from this, one can find the 4d Planck scale $M_{pl}$ in terms of $M_5$. In our string theoretical set up $M_{pl}$ can be obtained most easily in the string frame and it turns out to be $M^2_{pl} \sim (1/\sqrt{-\lambda})(M^3_5 / g^2_s ) (1-e^{-\sqrt{-\lambda} \pi R_c})$. So if we assume $\sqrt{-\lambda} R_c \sim O(1)$, then using $1/\sqrt{-\lambda} \sim R_c$ we will have $M^2_{pl}\sim M^3_5 R_c /g^2_s$. This equation may be regarded as a string theoretical generalization of the conventional equation $M^2_{pl} \sim M^3_5 R_c$ \cite{8}, but in our case $R_c$ need not be so large to obtain $M_{pl} \sim 10^{19}GeV$ in contrast to the case of \cite{8}. Indeed, there is an interesting way of viewing this equation. If we take the hierarchy assumption that electroweak scale $m_{EW}$ is the only fundamental short distance scale in nature (i.e., if we assume that $M_5 \sim M_s \sim TeV$ and $R_c \sim TeV^{-1}$), the above equation is satisfied if we take $g_s \sim 10^{-16}$, which is just the realistic decoupling limit of LST \cite{32}, showing that this configuration also accords with the hierarchy assumption.

In the case  $\sqrt{-\lambda} R_c \ll 1$, we still get the same result as above ; $M^2_{pl} \sim M^3_5 R_c / g^2_s$, so the same story goes on. Namely if we take $M_5 \sim TeV$ and $R_c \sim TeV^{-1}$, then we get $g_s \sim 10^{-16}$. But in this case the condition $M_5 \sim M_s$, which was mentioned above in relation with subcritical strings, is not respected. Finally if $\sqrt{-\lambda} R_c \rightarrow \infty$, we get $M^2_{pl} \sim M^6_5 / M^4_s g^2_s$ by using $\sqrt{-\lambda} \sim M^4_s /M^3_5$. In this case we can think of two possibilities. The first one is the case $\sqrt{-\lambda} \rightarrow \infty$, and therefore $M_s \gg M_5$. In this case $R_c$ need not be very large, but $g_s \rightarrow 0$ is not guaranteed (also the condition $M_s \sim M_5$ is not respected) and this case may be irrelevant to our discussion. But if $\sqrt{-\lambda}/M_s \sim O(1)$ and therefore $M_s \sim M_5$, $g_s$ is again $g_s \sim 10^{-16}$ upon taking $M_s \sim M_5 \sim TeV$. But in this case the size of $S_1$ must be very large, i.e., $R_c \rightarrow \infty$.

\vskip 1cm
\begin{center}
{\large \bf Acknowledgement}
\end{center}
This work was supported by the National Research Foundation of Korea (NRF) - Grant funded by the Korean Government (353-2009-2-C00045).

\end{document}